\documentclass[preprint]{elsarticle}
\usepackage{bm}% bold math
\usepackage{graphicx}
\usepackage{epsfig}
\journal{Nuclear Physics A}
\begin{document}
\begin{frontmatter}
\title{Measuring the Temperature of Hot Nuclear Fragments} 
\author{
S. Wuenschel$^{ab}$, A. Bonasera$^{bc}$, L.W. May$^{ab}$,  G.A. Souliotis$^{b,d}$, R.Tripathi$^{b}$, 
 S. Galanopoulos $^{b}\footnote{ Present address: Hellenic Army Academy, Department of Physical Sciences \& Applications, Athens, Greece} $, Z. Kohley$^{ab}$, 
 K. Hagel$^{b}$, D.V. Shetty$^{b}\footnote{Present address: Physics Department, Western Michigan University, Kalamazoo, 49008, MI, USA }$, K. Huseman$^{b}$,
  S.N. Soisson$^{ab}$, B.C. Stein$^{ab}$and S.J. Yennello$^{ab}\footnote{E-mail address: yennello@comp.tamu.edu}$
}

%\affiliation{
\address{
a) Chemistry Department, Texas A\&M University, College Station, 77843, TX, USA\\
b) Cyclotron Institute, Texas A\&M University, College Station, 77843, TX, USA\\
c)Laboratori Nazionali del Sud-INFN, v. S.Sofia 64, 95123 Catania, Italy\\
d)Laboratory of Physical Chemistry, Department of Chemistry, National and Kapodistrian University of Athens, Athens, Greece\\
}
\begin{abstract}
A new  thermometer based on fragment momentum fluctuations is presented.  This thermometer exhibited residual contamination from the collective motion of the fragments along the beam axis.
  For this reason, the transverse direction has been explored.  Additionally, a mass dependence was observed for this thermometer.  This mass dependence may be the result of the  Fermi momentum of nucleons or the different properties 
  of the fragments (binding energy, spin etc..) which might be more sensitive to different densities and temperatures of the exploding fragments. We expect some of these aspects to be smaller for
   protons (and/or neutrons); consequently, the proton transverse momentum fluctuations were used to investigate the temperature dependence of the source.  
\end{abstract}

\end {frontmatter}
%\maketitle
\section {Introduction}
For many years, various thermometers have been used to expand the experimental understanding of nuclear systems.  These studies have often been motivated by the desire to define the proposed nuclear liquid-gas phase transition~\cite{Kelic2006,Borderie2007,Bonasera08}.  
A broad array of caloric curves have been obtained allowing a better understanding of the nuclear limiting temperature and its dependence on source excitation energy as well as source size~\cite{Natowitz2002}. 
 Additionally, nuclear thermometry has provided a bridging connection between isoscaling~\cite{Tsang2001a,Tsang2001b,Xu2000} and the symmetry coefficient of the nuclear equation of state~\cite{LeFevre2005}.  

Several thermometers have been used extensively in nuclear studies.  In many studies, the nuclear temperature has been obtained from energy spectra through moving source 
fitting~\cite{Wada1989,Hirsch1984,Odeh2000}. However, this thermometer is known to exhibit non-thermal and collective behavior~\cite{Bauer1995}.
  Alternatively, temperature may be obtained through isotopic thermometers~\cite{albergo}.  The double isotope thermometer has provided much data~\cite{Viola2006,Tsang1997,Hsi1994,Trautmann2007},
   however the temperatures derived are complicated by model dependent secondary decay corrections~\cite{Trautmann2007,Wada1997}.  Ratios of excited states may also be used to obtain temperature for low to mid excitation energies~\cite{Borderie2007}.
  %In an effort to better understand the nuclear temperature, a new thermometer based on momentum fluctuations is derived and tested experimentally.  

\section {Experimental}
In this work, we seek to measure the temperature of reconstructed quasi-projectiles from the reactions of $^{86,78}$Kr+$^{64,58}$Ni at 35MeV/nucleon collected with the NIMROD-ISiS array~\cite{Wuenschel2009, Wuenschel2009a} at Texas A$\&$M University. 
The NIMROD-ISiS array is a $4\pi$ charged particle array housed inside the TAMU Neutron Ball.  This experimental hardware configuration provides free neutron data  in conjunction with isotopically resolved charged particle event reconstruction. 
The data presented in this paper was extracted from particles detected in the silicon detectors in rings 2-9 and the CsI detectors in rings 2-11. 

\begin{figure}[t]
\leavevmode
\includegraphics[width=0.5\textwidth,keepaspectratio=true,angle=0]{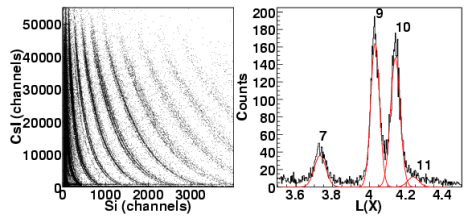}
\caption{ Left: Raw data from $\Delta$E-E with Si-CsI detectors. Right: Projected, linearized data for Z=4 isotopes. The x-axis is L$_{X}$, which is the linearized value. (color online)}
\label{fig3}
\end{figure}

There are three possible sources of fragment charge and mass identification in NIMROD-ISiS.  Pulse shape discrimination of the CsI fast versus slow light output was used for Z=1,2.  The $\Delta$E-E method is used for $Z\geq3$ on Si-Si and Si-CsI. 
 The particle identification was done through linearization of the raw data~\cite{Wuenschel2009a} to remove the non-trivial curvature due to energy deposition characteristics of the detection methods (see left panel of Fig.~\ref{fig3}). 
 The linearization utilized lines carefully chosen to follow the strongest isotope of each element.  The data were then straightened using a calculation of the distance between the data point and two closest chosen lines. 

After linearization, data were projected onto an axis, L(x), producing quasi-Gaussian peaks. The isotopic peaks in the projected distributions were fitted with Gaussian functions. 
The fragment charge was assigned by limits placed on the linearization x-axis value (L$_{X}$). The mass of each particle was assigned by determining the probability of the particle belonging to a given isotope.
 This probability (P$_A$) was calculated by comparing the value of the the isotopic Gaussian functions at the L$_X$ value of the particle
\begin{equation}
P_{A} = \frac{G_{A}(L_{X})}{\sum _{i}G_{i}(L_{X})}
\label{GausProb}
\end{equation}
where G$_A$ is the fitted Gaussian to the selected isotope which is compared to the summation over all Gaussian functions (G$_i$) defined for the element. For this data, a  mass was defined only if the P$_A$ was $\ge 0.75$.
 This method of fitting the linearized data with Gaussian functions provided the ability to estimate the average contamination between neighboring isotopes. The average contamination in the yield of a given isotope as defined here was calculated to be
  $\approx$ 5$\%$ across all reaction systems and all detectors.  Example linearized projections of Be fragments are plotted in the right panel of Fig.~\ref{fig3}.

The quasi-projectile source was selected by means of several event-by-event cuts on the experimental data.  The first cut required that the sum of the collected charged fragments ($\sum Z$) for the event equal a minimum of Z=30.  
The fragments in an accepted event were then cut on the longitudinal velocity relative to the largest fragment~\cite{Steckmeyer2001}.  This cut varied with the fragment Z.  The fragments retained for Z=1, Z=2, and Z$\geq$3 had 
to have longitudinal velocities within the range of $\pm 65\%, 60\%, 40\%$, respectively, of the  largest fragment longitudinal velocity.  The sum of the charges of the collected and accepted fragments was again constrained to be 
in the range Z=30-34.  Limits were placed on the deformation of the source by means of a cut on the quadrupole moment of the momentum distribution~\cite{Wuenschel2009b}.  

This method of source definition was compared to events and fragments generated by the HIPSE-SIMON code~\cite{Lacroix2004}.  
The initial $\sum Z$ cut eliminates a significant number of incomplete events and events from other sources.  This cut, however, 
was not sufficient to eliminate all fragments resulting from mid-velocity or pre-equillibrium sources.  The fragment velocity cut
 successfully helped minimize fragments from non-projectile-like sources.  The final cut on the deformation provided a reasonable level of isotropy in the selected events.  

After the quasi-projectile source was defined, the event excitation energy was calculated using: 
\begin{equation}
E_{source}^{\ast} = \sum_{i}^{M_{cp}} K_{cp}(i) + M_{n}\left<K_{n}\right>-Q_{gg}.
\label{Excite}
\end{equation}
The excitation energy (E$^{\ast}_{source}$) was defined as the sum over the accepted particles center of mass kinetic energy (K$_{cp}$ and K$_{n}$) minus the reaction Q-value.
 The average kinetic energy of the neutrons was calculated as the proton average kinetic energy with a correction for the Coulomb barrier energy~\cite{Dore2000}. 

Proper calculation of the excitation energy should include the neutron and gamma energy corrections. The Neutron Ball provides an experimental neutron multiplicity per event that must be corrected for background,
 neutron source, and efficiency. The multiplicity of free neutrons was extracted from the experimental quasiprojectile (Q$_p$) data using:
\begin{equation}
\label{NN}
M_{QP}=\frac{ M_{exp}-M_{bkg} }{ (\epsilon _{QP}+\frac{N_{T}}{N_{P}} \epsilon _{QT})(.7/.6)}.
\end{equation}
The multiplicity of neutrons assigned to the projectile source (M$_{QP}$) was calculated from the background (M$_{bkg}$) corrected experimental neutron multiplicity (M$_{exp}$). 
 and corrected using the relative efficiency of the Neutron Ball for detecting free neutrons emitted from a
  quasi-projectile ($\epsilon  _{QP}$) and quasi-target ($\epsilon  _{QT}$) sources for this reaction. The free neutron correction also accounted for the total neutron contributions from both the target
   (N$_{T}$) and projectile (N$_{P}$) nuclei.  The efficiencies were extracted from tagged neutrons generated by the HIPSE-SIMON code and the GEANT3~\cite{GEANT} simulation of the detector~\cite{Wuenschel2009}.  
    Simulation and filtering provided $ \epsilon  _{QP}$=0.65 and $\epsilon  _{QT}$=0.40. The Neutron Ball was calibrated with a $^{252}$Cf source to 70$\%$ efficiency for this experiment. 
    The correction factor in the denominator accounts for the experimental calibration of the Neutron Ball to 70$\%$ efficiency rather than the 60$\%$ efficiency obtained from the GEANT3 simulation of the detector.  
    Event by event, the M$_{QP}$ provided an estimate of the free neutrons emitted from the quasi-projectile.

The energy carried away by gammas is generally considered to be small~\cite{Viola2006,Lefort2001}.  In an effort to account for gamma energy, the ISiS collaboration calculated $E_{\gamma}=1 MeV(M_{Z\geq3})$~\cite{Lefort2001}.
  However, this energy has not been constrained experimentally~\cite{ViolaPC}.  Using this formula the gamma energy is $\leq$ 1\% of the total excitation energy of these events.  The gamma energy correction is not included in 
  calorimetry of this data as it is small compared to that of the excitation energy and is poorly constrained.  
 \section{Temperature determination with the momentum fluctuation thermometer}
\begin{figure}[t]
\begin{center}
\epsfig{file =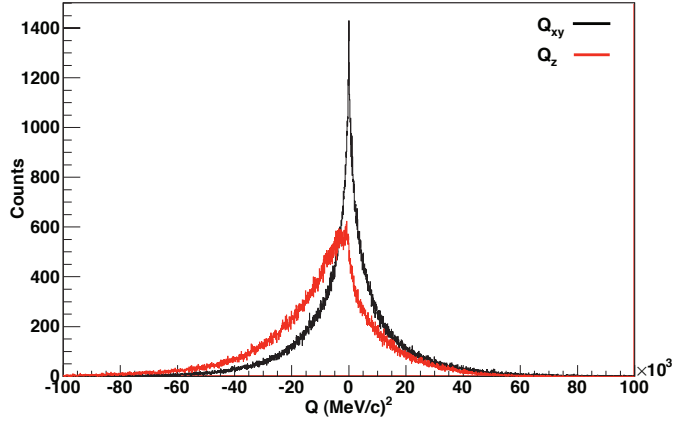,width=9.5cm,angle=0}
\end{center}
\caption{Q$_{z}$ and Q$_{x,y}$, Q distributions extracted from equation ~\ref{T_Q}  for the longitudinal (Q$_z$) and transverse (Q$_{xy}$) calculated using all fragments. (color online). }
\label{Qdist}
\end{figure}

A new method of calculating the system temperature can be derived from the momentum fluctuations of particles in the center of mass frame of the fragmenting source.  In this case, the frame is that of the reconstructed quasi-projectile. 
 The momentum is constructed for each particle in the projectile frame center of mass for each axis.  The momenta are then compared using Eq.~\ref{T_Q}:
\begin{equation}
\begin{tabular}{l c l}
Q$_{x}$ &= &2P$_{x}^{2}-P_{y}^{2}-P_{z}^{2}$\\
Q$_{y}$ &= &2P$_{y}^{2}-P_{x}^{2}-P_{z}^{2}$\\
Q$_{z}$ &= &2P$_{z}^{2}-P_{y}^{2}-P_{x}^{2}$ = 2P$_{z}^{2}-P_{T}^{2}$\\
Q$_{xy}$ &= &P$_{x}^{2}-P_{y}^{2}$\\
\label{T_Q}
\end{tabular}
\end{equation}
where P$_{x}$, P$_{y}$, and P$_{z}$ are the momenta in the X,Y and Z-axis respectively of each particle.   The Q distributions extracted from equation ~\ref{T_Q}  for the longitudinal (Q$_z$) and transverse (Q$_{xy}$) are shown in figure ~\ref{Qdist}.    In an ideal, spherical, fragmenting source the sum of Q for all fragments in the event in the center of mass of that event should 
be a $\delta$ function at zero.  However, in a real system fluctuations occur within a class of events.  These fluctuations convert the momentum $\delta$ function into a distribution that can be characterized by a mean and width.
 In an equilibrated system, the mean should still equal zero.  

The variance ($\sigma^{2}$) may be obtained from the Q$_{z}$ distribution through
\begin{equation}
\sigma^{2} = \left < Q_z^{2} \right > - \left < Q_z \right >^{2}
\label{Var}
\end{equation}
where Q$_z$ is the quadrupole moment (Eq.~\ref{T_Q}).  If the mean equals zero, the second term cancels.  Taking the first term as
\begin{equation}
 \left <Q_z^{2}\right > =
 % \int d^{3} p \left ( 2P^{2}_{Z}  - P^{2}_{X} - P^{2}_{Y}\right )^{2}f(p) =
  \int d^{3}p \left ( 2P^{2}_{Z} - P^{2}_{T} \right )^{2} f(p)
\end{equation}
and assuming a Maxwellian distribution of the momentum yields
\begin{equation}
 \left <Q_z^{2}\right > = \frac{1}{\left (2\pi mT \right ) ^{3/2}} \int d^{3}p \left ( 4P^{4}_{Z} - 4P^{2}_{Z}P^{2}_{T} + P^{4}_{T} \right ) e^{ - \frac{P^{2}_{Z}  + P^{2}_{X} + P^{2}_{Y}}{2mT} }.
 \label{Q2}
\end{equation}

A Gaussian integral is used
\begin{equation}
I_{n}(a)= \int _{0}^{\infty} x^{n}e^{-ax^{2}}dx
\label{GausIntegral}
\end{equation}
and the results that are of interest to this derivation are:
\begin{equation}
I_{0}(a)= \frac{1}{2}\sqrt{\frac{\pi}{a}}
\end{equation}
\begin{equation}
I_{2}(a)= \frac{1}{4a}\sqrt{\frac{\pi}{a}}
\end{equation}
\begin{equation}
I_{4}(a)= \frac{3}{8a^{2}}\sqrt{\frac{\pi}{a}}.
\end{equation}

It is important to note that these integrals are only for 0 to $\infty$ and therefore should be doubled for -$\infty$ to $\infty$.
After derivation, the relation between variance and temperature becomes:

\begin{equation}
\sigma ^{2} = 4\left ( \frac{3}{4a^{2}} \right ) - 4\left ( \frac{1}{2a^{2}} \right ) + \frac{2}{a^{2}} = 12m^{2}T^{2}.
\end{equation}

where
\begin{equation}
a=\frac{1}{2mT}
\end{equation}

For a single fragment type from a nuclear multi-fragmentation:
\begin{equation}
\sigma ^{2}  = 12A^{2}m_{0}^{2}T^{2}
\end{equation}
where m$_{0}$ is the mass of a nucleon and A is the mass number of the fragment.   For multiple particle types, the formula becomes:
\begin{equation}
\sigma ^{2} = 12m_{0}^{2}T^{2}\sum_{i} \left(\zeta_{i}A_{i}\right)^{2}
\label{T_many}
\end{equation}
where $\zeta$ is the concentration of the particle in question.  A similar derivation can be carried out using Q$_x$ and Q$_y$ defined in Eq.~\ref{T_Q}.

Because of the velocity  source cuts, the Z-axis does not exhibit the same momentum distributions as are seen in the X,Y-axes.  
Therefore, the temperature was re-derived using only the P$_{x}$ and P$_{y}$ momentum and employing the same Maxwellian distribution and Gaussian integral assumptions.
  Here Q$^{2}_{xy}$ is defined as
\begin{equation}
 Q^{2}_{xy} = \int d^{2} p \left ( P^{4}_{X}  - 2P^{2}_{X}P^{2}_{Y} + P^{4}_{Y}\right )f(p)
\end{equation}
and using the Maxwellian distribution assumption gives:
\begin{equation}
\sigma ^{2} = 4m^{2}T^{2}.
\end{equation}

\begin{figure}[t]
\begin{center}
\epsfig{file =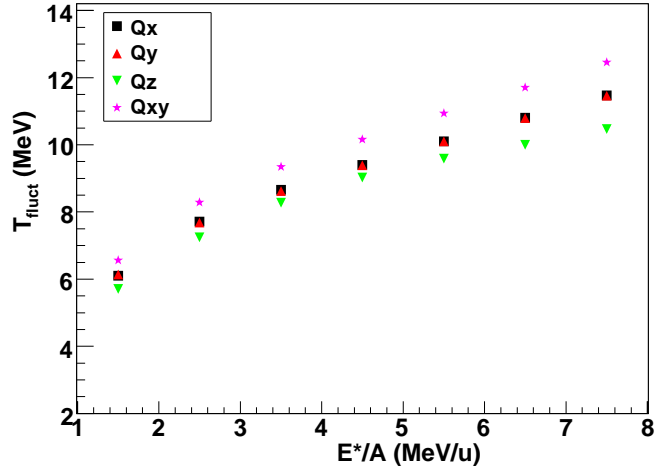,width=9.5cm,angle=0}
\end{center}
\caption{Q$_{x}$, Q$_{y}$, Q$_{z}$  and Q$_{xy}$ (Eqs.~\ref{T_Q} and~\ref{T_XY}) temperatures (Eqs.~\ref{T_many} and~\ref{T_XYmany}) calculated using all fragments as a function of E$^{\ast}$/A of the source event. (color online).  }
\label{T_Qs}
\end{figure}

For a single fragment type from a nuclear multi-fragmentation:
\begin{equation}
\sigma ^{2}  = 4A^{2}m_{0}^{2}T^{2}
\label{T_XY}
\end{equation}
where m$_{0}$ is the mass of a nucleon and A is the mass number of the fragment. For multiple fragment types, 
\begin{equation}
\sigma ^{2} = 4m_{0}^{2}T^{2}\sum_{i} \left(\zeta_{i}A_{i}\right)^{2}
\label{T_XYmany}
\end{equation} 
where $\zeta$ is again the concentration of a given particle type. The temperature is now linked to the variance through Eqs.~\ref{T_many}  and ~\ref{T_XYmany}. %and the caloric curve obtained using the Q$_{xy}$ fluctuations is plotted in Fig.~\ref{T_Qs}, see also Fig.~\ref{Qdist}. 
The temperature evaluated from all fragments assigned to the quasi-projectile fragmentation event  is shown in Fig.~\ref{T_Qs} for the quadrupole oriented along various axes (Q$_z$,Q$_x$,Q$_y$) as well as the transverse quadrupole (Q$_{xy}$) (Eq.~\ref{T_Q}). 
These quantities should be exactly the same in a thermally equilibrated system, any deviation from thermal equilibrium will be disclosed from the differences in the quantity Q calculated
along different axes.  The statistical error bars are smaller than the size of the points in Fig.~\ref{T_Qs}.  The Q$_{x}$, Q$_{y}$ thermometers are identical within errors and the Q$_{z}$ thermometer is slightly lower due to the longitudinal velocity cuts imposed as part of the quasiprojectile source selection. 
The transverse thermometer based on Q$_{xy}$ provides a higher source temperature across all excitation energies than was obtained using the Q$_{x}$, Q$_{y}$, or Q$_{z}$ thermometers.  This difference in temperature is the result of eliminating the effect due to the velocity cuts along the Z-axis. 
Hence it is imperative to investigate the temperature extracted from the transverse velocity of the fragments.

%In order to have events that were not contaminated with excess momentum in the beam direction fragments had to have a longitudinal velocity similar to the largest fragment as discussed earlier.  However these cuts artificially lower the derived temperatures.  

% This behavior is likely the result of residual collective motion along the beam axis. 
 %Thus, when this axis is used to calculate Q$_{x}$, Q$_{y}$, or Q$_{z}$ the distributions are contaminated by the remaining collective behavior. It is interesting
%to note that a residual collective motion results in a smaller 'apparent' temperature.  Thus it is important to eliminate as much as possible the residual collective motion from the event characterization.  As we will show, this has
%other interesting consequences in the determination of the caloric curve as discussed below.

The momentum distributions may also be widened by recoil during the source break up.  The recoil effect on the momentum Q distributions has been parameterized as
\begin{equation}
Q_{xy(0)}=Q_{xy(app)} \left( \frac{A-A_{f}}{A-1} \right)
\label{Recoil}
\end{equation}
where Q$_{xy(0)}$ is the recoil corrected value, Q$_{xy(app)}$ is the experimental value (Eq.~\ref{T_XYmany}), A is the mass of the source, and A$_{f}$ is the mass of the fragment being considered~\cite{Bauer1995}. 

\begin{figure}
\begin{center}
\epsfig{file =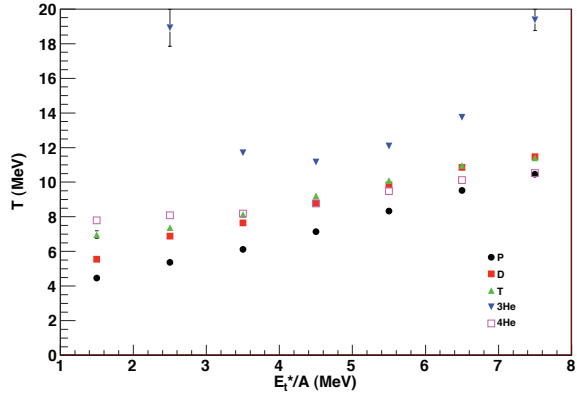,width=8.5cm,angle=0}
\end{center}
\caption{Apparent temperatures extracted from the Q$_{xy}$ distributions as a function of particle type. (color online)}
\label{SigmaDist}
\end{figure}

Combining all of the fragments together to create a Q distribution depends on the assumption that all of the fragments, regardless of mass or charge, have the same distribution width.  This implies a simultaneous, statistical emission of fragments. 
The apparent temperatures obtained from the recoil corrected Q$_{xy}$ distributions as a function of particle type and source excitation energy are plotted in Fig.~\ref{SigmaDist}.  Though the temperature increases with excitation energy regardless of particle type,
  there is a significant spread in measured temperature across the fragment types.

 A  significant mass dependence at constant Z may be seen in the $^{3,4}$He isotopes. The $^{3}$He yields a $\sim$35$\%$ higher temperature than the $^{4}$He across all E$^{\ast}$/A.
   The $^{4}$He apparent temperatures show a $\sim$15$\%$ increase across E$^{\ast}$/A= 1.5--7.5.  The $^{3}$He temperatures, on the other hand, increase by $\sim$50$\%$ across the same E$^{\ast}$/A region.  It is possible that some of the mass dependence of the apparent temperature is the result of differential fragment emission time.    It is known that $^{3}$He is emitted early in the fragmentation process.  Conversely, $^{4}$He is emitted throughout the de-excitation cascade~\cite{Viola1999}.  
The difference in measured temperature seen in the A=3 isotopes may indicate a Coulomb contribution to the momentum distributions widths. On the other hand the results for $^2$H and $^4$He isotopes rules out a strong Coulomb dependence of T.   Other authors results [3] suggest that Coulomb is not so important in our beam  energy region.

In addition to the above mentioned factors, the temperature obtained with this thermometer could be a combination of thermal energy and Fermi momentum in the detected fragments~\cite{Bauer1995}.  
Following the paper of Bauer, the measured fragments exhibit a momentum distribution resulting not only from thermal sources, but also from the Fermi momentum of the component nucleons with the fragment.  
This effect increases as a function of fragment size until a limiting value is observed. This correction is only meaningful for fragments with A$\geq$2.  Thus, the momentum distributions of protons are free of this complication.  

\begin{figure}[t]
\begin{center}
\epsfig{file =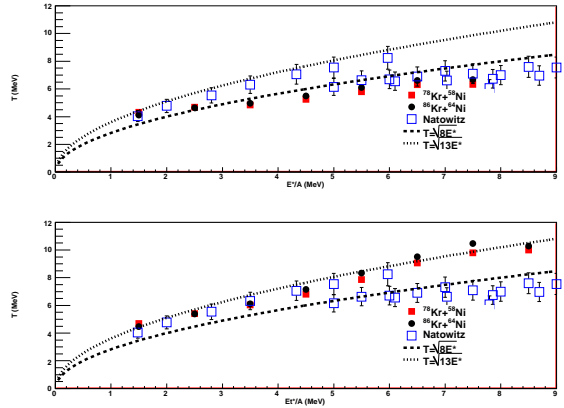,width=8.5cm,angle=0}
\end{center}
\caption{Q$_{z}$ (Eq.~\ref{T_Q}) temperatures derived from proton momentum fluctuations as a function of E$^{\ast}$/A of the source event. For reference the caloric curve for A=60-100
 from the Natowitz compilation~\cite{Natowitz2002} as well as two Fermi Gas ( T = $\sqrt{aE^{\ast}/A}$ with a = 8, 13 ) curves are plotted(Top panel). (Bottom panel) same as above but for the
 corresponding 'transverse' quantities. (color online)}
\label{Protons}
\end{figure}

The caloric curve derived from the proton momentum fluctuation widths of the $^{86}$Kr+$^{64}$Ni system is plotted in Fig.~\ref{Protons} (top panel).  
The fluctuations for this thermometer include the Z-axis, eq.(16), thus include collective effects if any.
  For reference, the compilation of Natowitz \textit{et al.}~\cite{Natowitz2002} and two Fermi Gas curves ( T = $\sqrt{E^{\ast}/a}$ with a = A/8, A/13 MeV$^{-1}$) are plotted. 
 As shown in this figure, the protons produce temperatures similar to what is obtained from a Fermi Gas at low E$^{\ast}$/A indicating that this thermometer does indeed yield reasonable temperature values. 
  The errors plotted in Fig.~\ref{Protons} for the Y-axis are estimated systematic errors of 0.5 MeV.
  
As discussed above the quadrupole fluctuations displayed a nonequilibrium effect still present along the beam direction.  We expect these effects to be present in the determination of the excitation energy as well. To a first approximation we can correct for this 
nonequilibrium effect by replacing the kinetic energy of the fragments in eq.(2) by its transverse component. i.e. : $K_{cp}\rightarrow 3/2 K_{Tcp}$, similarly for neutrons.
 The excitation energy determined from the transverse kinetic energy of the fragments is approximately 0.3to 0.5 MeV larger than that which is determined by the total kinetic energy, see Fig.~\ref{Protons} (bottom panel).
 As we see in the figure, the difference when the transverse energy and the 'transverse' temperature are used becomes quite important for increasing excitation energies where collective effects become more important.  Interestingly
  the 'caloric curve' obtained without collective effects shows a back bending  similar to what was predicted by D. Gross long ago\cite{gross2}.  This might be a signature for a phase transition as discussed in \cite{gross97,gross2},
  however a more detailed study of this feature is beyond the goal of this paper and will be discussed in more detail in a future publication.
  
 When the collective motion is removed the transverse temperature rises from approximately 4 MeV to 10 MeV over this excitation energy range.  Slope "temperatures" were also extracted from this data set by fitting the kinetic energy spectra\cite {moretto}  of the fragments. Both the quadrupole temperature and the slope temperature are derived from the kinetic energy.  However the quadrupole temperature should be less susceptible to secondary decay and non-equilibrium effects.  
 Since the velocity cuts imposed to minimize the effect of the non-equilibrated motion in the beam direction mainly affect the Z=1 and Z=2 particles, the slope temperatures extracted from the fits to the kinetic energy spectra for Z$\ge$3 should not be impacted.
  The values of the slope temperature were consistently approximately 3 MeV higher than those extracted from the quadrupole temperature.
As already mentioned the fragment kinetic energy spectra are a result of both thermal energy and Fermi momentum\cite{Bauer1995}.  Thus this comparison may contain important information about the contribution from the Fermi momentum and will be the subject of future works.
  
  We expect that at low T or excitation energy the Fermi gas formula $E^*=aT^2$ should be valid, see Fig.5.  Thus it is 
convenient to plot the quantity $E^*/aT^2$ versus T to see any deviation from a pure Fermi gas.  
In Fig (6) we plot the reduced excitation energy versus temperature as derived from the transverse kinetic energy of the fragments (full squares) as well as the total kinetic energy of the fragments (open squares).  As seen in the figure at small T
the two results are identical within the error bars.  For increasing T the two results differ much demonstrating the increased influence of the non-equilibrium effects with increasing T or excitation energy.  The back bending is now more pronounced 
compared to fig. 5.

The values obtained from the transverse energy have been normalized to 1 at the lowest measured temperature using a level density parameter $a_0=A/13.3$ MeV$^{-1}$.
  We notice that such a quantity is density dependent i.e. $a=a_0(\rho_0/\rho)^{2/3}$.  
From this relation, assuming that the nuclear Fermi gas is reaching different densities at different temperatures
\cite{bertsch,natowitz,viola}, we can deduce a value of density for each measured temperature.
 %From this result 
%we can deduce that the system might had expanded to a density $\rho/\rho_0=1/3.2$ where $\rho_0$ is the density of the system at the lowest measured temperature.  
\begin{figure}[t]
\begin{center}
\epsfig{file =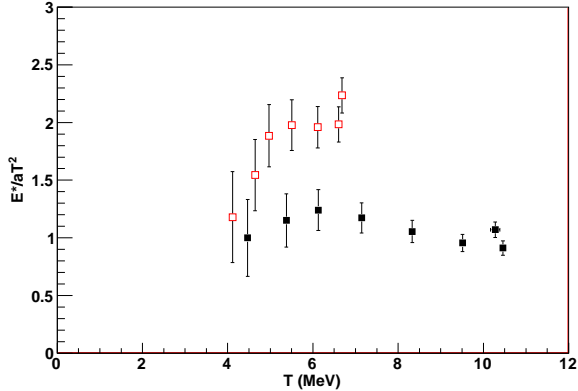,width=8.5cm,angle=0}
\end{center}
\caption{$E^*/aT^2$ vs T  derived from the transverse kinetic energy of the fragments (full squares) and the total kinetic energy (open squares) for the $^{86}Kr+^{64}Ni$ system.   (color online)}
\label{E*T}
\end{figure} 
We can easily calculate the density at each temperature by noticing that:
\begin{equation}
\frac{\rho}{\rho_0}=(\frac{a_0T^2}{E^*})^{3/2}
\label{density}
\end{equation}
The validity of the Fermi gas model is related to the ratio of the temperature to the Fermi energy $T/\epsilon(\rho)_F$ .  If the density decreases also the Fermi energy decreases as: $\epsilon(\rho)_F=36 (\rho/\rho_0)^{2/3}(MeV)$.
Where $36 MeV$ is the Fermi energy of the nucleus ground state and we are assuming to be the same value at the smallest measured temperature.  
  Since we can estimate the change of density for each temperature we can also obtain a value for the Fermi energy.  In figure (7) we plot the reduced density versus 
the reduced T ( i.e. divided by the corresponding Fermi energy).   As we see in the figure the values of the reduced temperature are well below 1 which implies that our Fermi gas approximation is valid and the system is 
a quantum Fermi system even at the highest measured temperatures.  This is at variance with many authors which assume that the system is classical at the highest measured temperature\cite{bon00}. More important,
the result without collective effects suggests that the density is roughly constant in the range of temperatures considered at variance with the analysis which includes collective effects.  The latter case would suggest
an expanding system at all temperatures while the previous case enforces the concept of freeze out volume.  If this conjecture will be confirmed by future analysis we could study the nuclear equation of state at (almost) constant density.
\begin{figure}[t]
\begin{center}
\epsfig{file =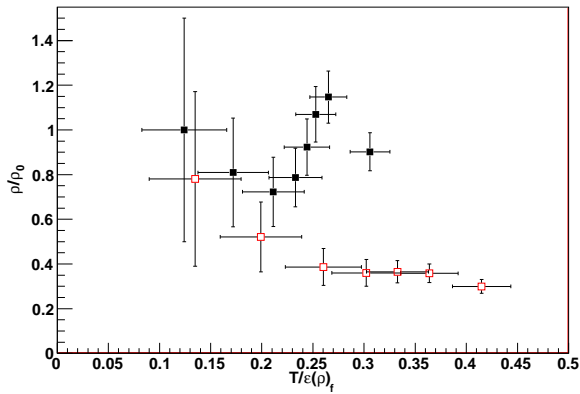,width=8.5cm,angle=0}
\end{center}
\caption{Reduced density vs reduced temperature derived for a Fermi gas from the transverse kinetic energy of the fragments (full squares) and the total kinetic energy (open squares) for the $^{86}Kr+^{64}Ni$ system.   (color online)}
\label{rhoT}
\end{figure}

If the picture of an expanding system with increasing temperature is correct, a caloric curve as the one of figures (5) and (6) might be misleading.  A different physical interpretation of the 
caloric curve could result if we plot opportunely normalized quantities as in figure(8).  
\begin{figure}[t]
\begin{center}
\epsfig{file =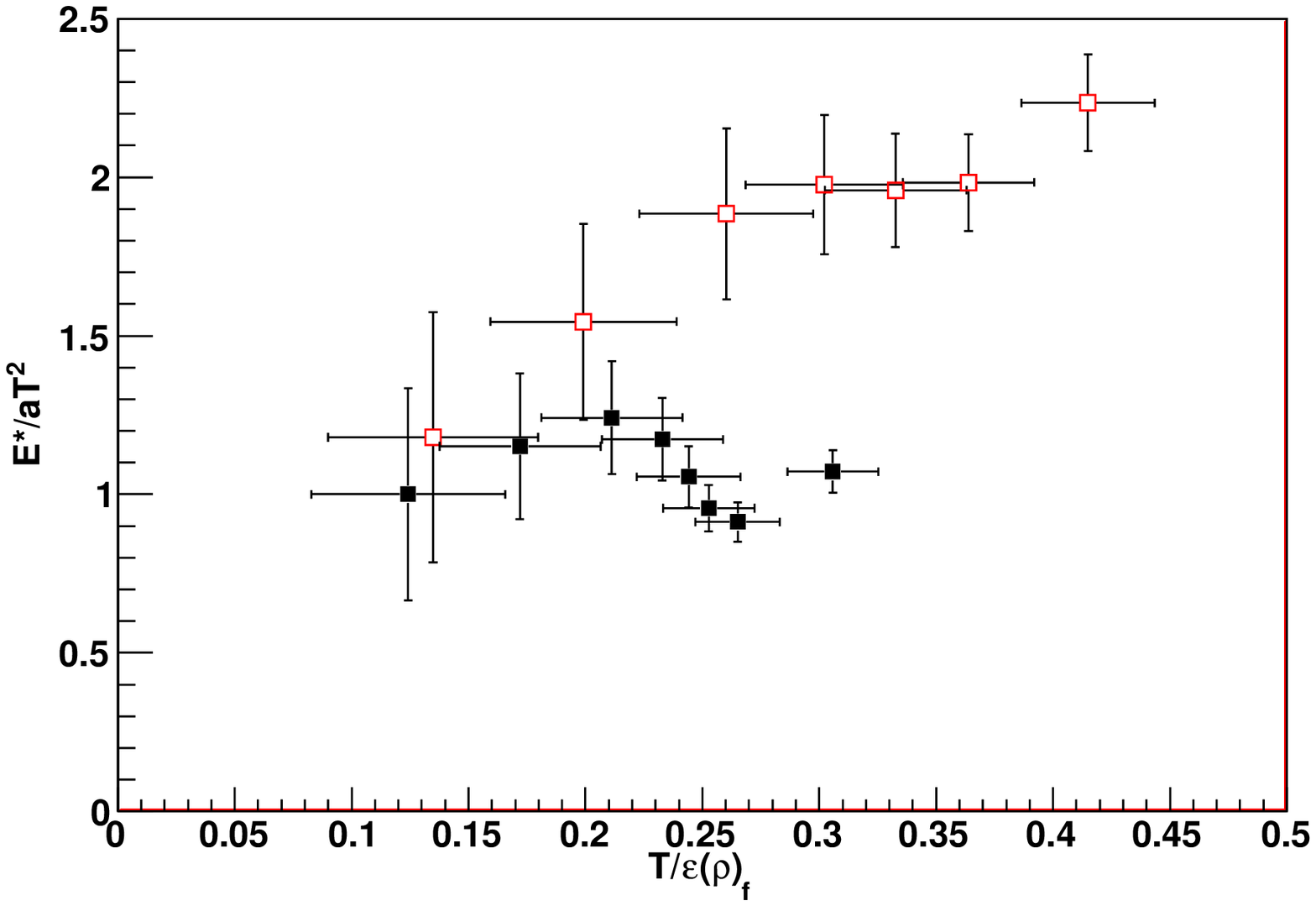,width=8.5cm,angle=0}
\end{center}
\caption{Caloric curve in reduced units for the $^{86}Kr+^{64}Ni$ system.  Full squares are from the transverse kinetic energy of the fragments and the open squares are using the total kinetic energy. (color online)}
\label{reduced}
\end{figure} 
In these units the 'apparent' saturation of the temperature at high excitation energies now disappears because of the changing density of the system contained in the Fermi energy and the level density
parameter.  In contrast the results with non-equilibrium effects removed display a more remarked back bending.  We stress that these results are derived assuming a Fermi gas relation and are for sake of discussion.
A more precise determination of the equation of state requires a more complete determination of the quantities at play, i.e. excitation energy, temperature and density (or pressure) of the system.  

%  Unfortunately the experimental error bars are too large to determine if a change of slope occurs at some reduced T.  It would be of course interesting to repeat this analysis for smaller and higher
%excitation energies than those included in this paper. 
\section {Concluding remarks}
Four momentum thermometers have been defined here (Eqs.~\ref{T_Q},~\ref{T_XYmany}).  The first three (Q$_{x}$, Q$_{y}$, and Q$_{z}$) should theoretically be degenerate.  However, the Z-axis was found to be truncated by the source selection.
  This affected all three of these thermometers.  To remove this characteristic along the Z-axis, the temperature was then derived using only the X and Y-axes. 
   This thermometer provides a slightly higher temperature reflecting the removal of the Z-axis with its incorrect width. The high value of T obtained from this thermometer may be a result of Fermi momentum in the fragments.  
   This effect is zero for protons.  The proton caloric curve is shown in Fig.~\ref{Protons}. Since a non-equilibrium effect is present in the temperature determination, we have also suggested a method to reduce
   the non-equilibrium effects in the determination of the excitation energies by using the transverse kinetic energy of the fragments.  The modified caloric curve thus obtained has been discussed in Fig.~\ref{Protons} demonstrating
   the increased importance of collective effects at high system temperatures (or excitation energies).  Assuming a Fermi gas relation for an expanding system we have been able to derive a density value 
   for each temperature.  From the density we can estimate the corresponding Fermi energy and demonstrated that the system is still a quantum Fermi system at the current  temperatures and densities.  
   We stress that the assumption of an expanding Fermi gas is suggested by some microscopic calculation \cite{bon00}, but it is at variance with some statistical models which assume a constant
   freeze-out density \cite{gross97}.  If a constant density is the correct picture then the best representation of the caloric curve would be the ones of Fig. 5 and Fig. 6 while if density is changing we should refer to Fig. 8.  In both
   cases we do not expect the densities to be dramatically different from those estimated in Fig.7 which implies that we are in a quantum regime at all temperatures.
   
   Which picture of fragmentation is correct will probably be debated both experimentally and theoretically in future years. To this debate we have to add the observation that our 
   proposed method gives different temperatures for different fragments.  The origin of this difference could be in the statistics of the fragments (i.e. Bose versus Fermi systems),
   in different Coulomb charges or time of creation including evaporation.  Other proposed thermometers also suffer from this isotope dependence of the temperature.  These features
   should be resolved in order to gain full understanding of an excited quantum nuclear system.
%\acknowledgements

\section{Acknowledgements}
The authors would like to thank Joe Natowtiz for stimulating discussions and the Cyclotron Institute for the excellent beam quality.  This work was supported in part by the Department of Energy under grant DE-FG03-93ER40773 and the Robert A Welch Foundation under grant A-1266.

\end{document}